\documentstyle[aas2pp4,eqsecnum,flushrt]{article}

\newcommand{\ii}{\'{\i}}
\newcommand{\etal}{{\em et~al.}}

\newcommand{\beq}{\begin{equation}}
\newcommand{\eeq}{\end{equation}}
\newcommand{\beqa}{\begin{eqnarray}}
\newcommand{\eeqa}{\end{eqnarray}}

\newcommand{\lsim}{\mathrel{\hbox{\rlap{\hbox{\lower3pt\hbox{$\sim$}}}\lower-2pt\hbox{$<$}}}}
\newcommand{\gsim}{\mathrel{\hbox{\rlap{\hbox{\lower3pt\hbox{$\sim$}}}\lower-2pt\hbox{$>$}}}}

\newcommand{\journ}[5]{
                      {#1},       
                      {#2}.       
                      {\it #3},   
                      {\bf #4},   
                      {#5}.       
                      }
\newcommand{\proceed}[6]{
                   {#1},        
                   {#2}.        
                in {\it  #3\/}, 
              eds. {#4},        
                   {#5},        
                p. {#6}.        
                        }
\accepted{March 21, 1997}
\paperid{E97-5021, bica.0113}
\cpright{AAS}{1997}
\slugcomment{Astrophysical Journal Letters. Accepted.}

\begin{document}

\title{Capture of field stars by globular clusters in dense bulge regions}

\author{Eduardo Bica, Horacio Dottori, Irapuan Rodrigues de Oliveira 
Filho\altaffilmark{1}}

\affil{Depart. de Astronomia, Instituto de F\'{\i}sica, UFRGS,  
        CP 15051, CEP 91501-907  Porto Alegre, RS, Brazil}

\author{Sergio Ortolani\altaffilmark{2}}
\affil{Universit\`a di Padova,  Dept. di Astronomia, Vicolo dell'Osservatorio 
5, I-35123, Padova, Italy}

\author{and}

\author{Beatriz Barbuy}
\affil{Universidade de S\~ao Paulo, IAG, Depto. de Astronomia, CP\,9638,
S\~ao Paulo\,01065, Brazil}
 
\altaffiltext{1}{ Brazilian CNPq fellow.}

\altaffiltext{2}{ Visiting astronomer at the European Southern Observatory, 
La Silla, Chile.}

\begin{abstract}
The recent detection of a double Red Giant Branch  in  the
optical color-magnitude diagram (CMD) of the bulge globular cluster HP1
(Ortolani et al. 1997), a more populated  metal-poor steep one corresponding
to the cluster itself,   and another metal-rich curved, led us  to explore in 
the present {\it Letter}  the possibility of capture of field stars by a globular 
cluster orbiting  in  dense bulge regions over several gigayears. Analytical arguments, 
as well as N-body calculationsLatex for a cluster model of $10^5 M_\odot$ 
in a bulge-like environment, suggest that 
a significant fraction of cluster stars may consist 
of captures. Metal-poor globular clusters in the inner bulge, like HP1, 
contrasting at least in $\Delta [Fe/H] = 1.0$\,dex with respect to 
the surrounding metal-rich   stars, are ideal probes to further test the capture scenario. In turn, if this
scenario is confirmed, the double RGB of HP1 could provide direct estimates 
of blanketing amounts, which is fundamental for the photometric calibration of metal-rich stellar populations.
\end{abstract}


\section{Introduction}
\label{intro}
Recently, Ortolani, Bica \& Barbuy (1997, hereafter OBB97) studied V vs 
V-I diagrams of 
the inner bulge globular cluster HP\,1, which  
revealed the presence of a double 
Red Giant Branch (RGB). One RGB is more populated and steep, 
and is  accompanied by a 
well-developed Blue Horizontal  Branch (HB), which  undoubtedly 
characterizes HP\,1 as a metal-poor globular cluster.  The other RGB,
observed in a  r$\leq\,23''$ extraction, 
is  curved and  extended, characteristic of a  nearly-solar metallicity  
stellar population, like that of the globular cluster NGC\,6553 
or the Baade
Window itself (Ortolani et al. 1995). OBB97 favored the interpretation of 
a bulge contamination 
over that of a merger of two clusters, to explain the secondary RGB
in HP\,1. In the present {\it Letter} we explore a scenario  where 
this contamination may correspond to physical captures of bulge stars.

Van den Bergh (1996) proposed the merger scenario to explain 
composite (blue and red) Horizontal Branches for some intermediate 
and low metallicity globular clusters in the Milky Way. 
For such mergers to occur, 
individual clusters must have low relative
velocities. This condition is matched in dwarf galaxies, 
where the clusters
could have merged in a first step inside the dwarf, and subsequently 
would have been accreted by our Galaxy. As pointed out by  van den Bergh (1996) the
exceptionally bright globular cluster M\,54 in the Sagittarius dwarf (Ibata, Gilmore \& Irwin 1994) might be such an example.

The merger scenario  would not be appropriate for the 
particular case
of HP\,1 since (i) the bulge velocity dispersion is very high, 
which makes unprobable mergers 
of globular clusters within the bulge,
and (ii) the stellar population of the secondary RGB in HP\,1
is exceedingly metal rich to have been originated in a dwarf galaxy.

In the present {\it Letter} we study,  both analytically and with N-body 
simulations, the possibility of    capture of bulge stars 
by a globular cluster, as
an alternative mechanism to explain the secondary RGB observed in the color-magnitude diagram (CMD)
of HP\,1.

\section{Analytical approach}
\label{anal}
We study the capture scenario by a globular
cluster orbiting inside the bulge, assuming
for the bulge total mass and mass distribution those predicted 
in the Galaxy model by
Hernquist(1993; see also, Hernquist 1990a). OBB97 derived for HP\,1 a distance of $\sim$1.3\,kpc 
from the Galactic center. For  this distance the cluster faces a star  
density of 0.33\,M$_\odot$/pc$^3$ (Hernquist 1993).
For the remaining cluster parameters we  consider a grid of values which
encompasses probable properties for a bulge globular  cluster.  
The rotation of the galactic bulge (Menzies 1990) implies that 
globular clusters in the bulge might present considerable  amounts of
streaming motion.  So we consider three possible
values for the cluster velocity (V$_c$): 50, 100 and 160\,km\,s$^{-1}$. For cluster 
diameters we test 10, 15 and 20\,pc, whereas for masses the values are 
10$^4$, 10$^5$ and 10$^6$\,M$_\odot$. For these ranges of sizes and velocities
the cluster encounters a total number of bulge stars in the range  
$1.3\times10^3\leq\,$N$_b\,\leq 7.1\times10^4$ during 1 Myr 
(assuming 1\,M$_\odot$ for bulge stars). Similarly, 
the  size and mass ranges
imply cluster escape velocities 3\,km\,s$^{-1}\leq$\,V$_e\,\leq 42$\,km\,s$^{-1}$. 
The cluster mass is the most important parameter determining the
escape velocity $V_e$, for the adopted grid.

The bulge velocity distribution is assumed to be isothermal (Binney\,\&\,Tremaine 1987) with a dispersion of 113\,km\,s$^{-1}$ 
(e.g. Sharples et al. 1991).  
The number of bulge stars capable to be captured by the cluster must 
satisfy the following velocity constraints:
$|V_c|\,-\,|V_e|\,\leq\,|V_s|\times\cos\theta\,\leq\,|V_c|\,+\,|V_e|$ 
and $|V_s|\times\sin\theta\,\leq\,|V_e|$
where $\theta$ is the angle between the star velocity ($\vec{V_s}$) and that of the cluster ($\vec{V_c}$).  Adopting  a typical mass of 1\,M$_\odot$ for bulge stars, 
the number  of stars  N$_s$ which are captured over 1\,Gyr 
by the cluster can be summarized as follows: 
(i) for a 10$^6$\,M$_\odot$ cluster with  escape velocity 
V$_e\,=\,34$\,km\,s$^{-1}$ and streaming velocity V$_c\,=\,50$\,km\,s$^{-1}$,  
N$_s\,=\,1.2\times10^5$,  whereas for V$_c\,=\,100\,$km\,s$^{-1}$,   
N$_s\,=\,4.1\times10^4$. (ii)  For a 10$^5$\,M$_\odot$ cluster with  
escape velocity 
V$_e\,=\,11$\,km\,s$^{-1}$ and streaming velocity V$_c\,=\,50$\,km\,s$^{-1}$,  
N$_s\,=\,5.1\times10^3$,  whereas for V$_c\,=\,100$\,km\,s$^{-1}$ the captures 
become  negligible. Finally, a 10$^4$\,M$_\odot$ cluster does not capture a
significant number of stars. This would set constraints on the mass of a 
cluster with a secondary RGB, like HP\,1.

\section{Numerical simulation}
\label{sec:simul}

\begin{figure*}[hp]
\vspace{18 cm}
\includegraphics{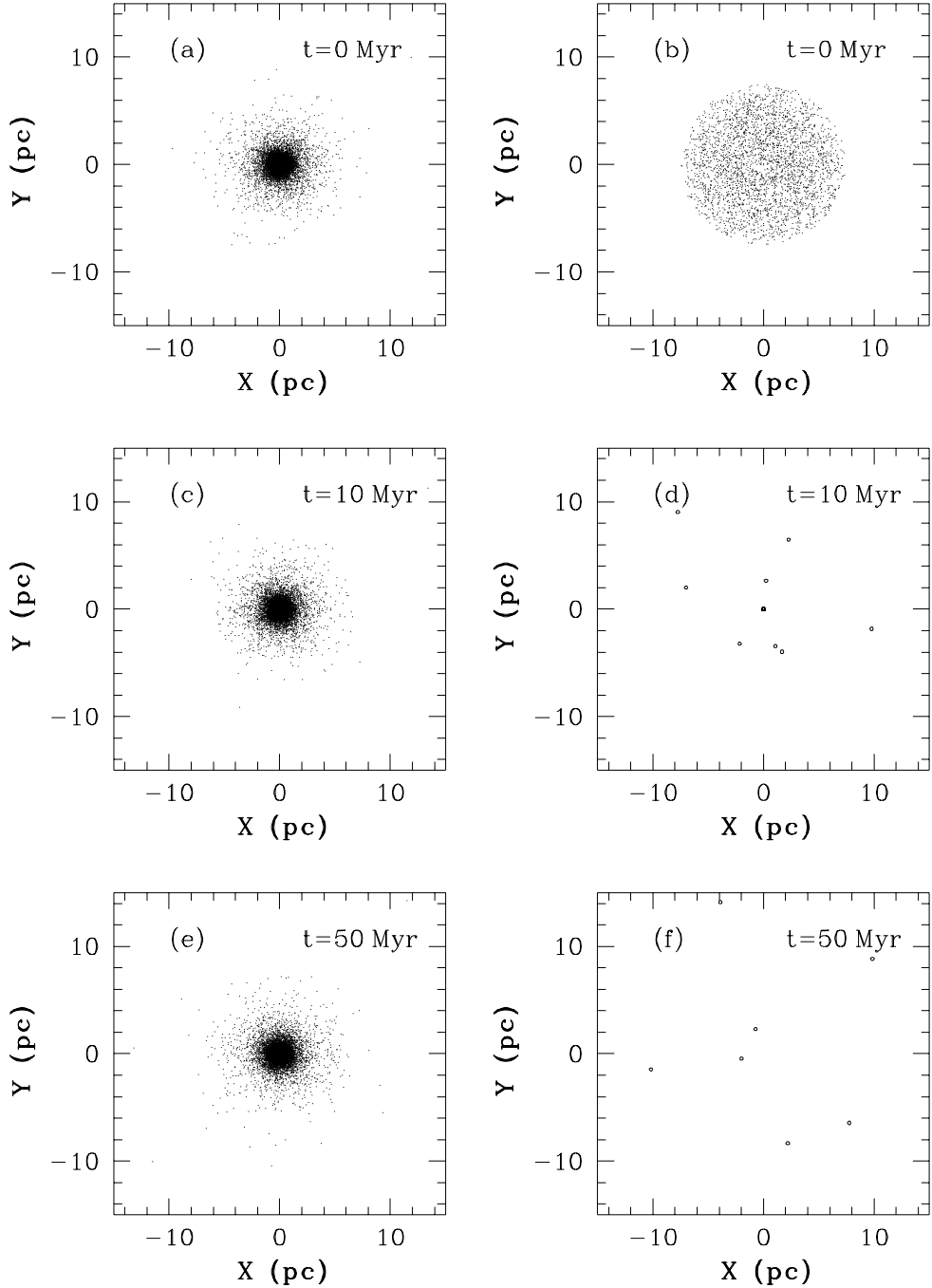}
\caption{Initial distribution of cluster and bulge 
particles encountered during 10\,Myr, respectively separated in (a) 
and (b) for clarity purposes.
(c) and (d) show respectively the cluster and  
nine  bulge particles  (90\,M$_\odot$)  captured  after 10\,Myr.
Finally, (e) and (f) test the cluster and capture
stability for further  40\,Myr.}
\label{fig:simul}
\end{figure*}                                                     

In order to further  check the capture scenario,
we performed a N-body numerical experiment. 
 We used a hierarchical 
tree algorithm (Barnes \& Hut 1986) in the CESUP-UFRGS Cray Y-MP2E computer. 
The  algorithm was optimized for vector architectures
(Hernquist 1987,1990b). The adopted tolerance 
parameter is 0.7, and the calculation includes quadrupole
terms.

The globular cluster 
was modeled with a Plummer polytrope with 10$^4$ particles
amounting a mass of 10$^5$M$_\odot$. The Plummer cluster cutoff 
diameter is 15\,pc
and the core diameter  is 2\,pc. The  cluster was left to evolve isolated 
during 1 relaxation time (2.8\,Myr) to check its stability.  Due to 
computational limitations on the number of particles in the simulation, 
we adopted for cluster and field stars 10\,M$_\odot$ particles.
More details on similar star cluster simulations are given in Rodrigues 
\etal (1994).

The adopted Galactic bulge density at the cluster location is that of 
the previous section. A cluster with a
streaming velocity of 50\,km\,s$^{-1}$ encounters a mass of bulge stars 
of  2948\,M$_\odot$\,Myr$^{-1}$. 
The bulge stars encountered by the cluster during  10\,Myr were placed 
in a homogeneous sphere  with size equal to that of the cluster, as shown in
Figs.~\ref{fig:simul}a,b.  The speed $|V_s|$ of a given star in the bulge 
sphere was randomly sampled from a Gaussian distribution with 
$\sigma=113$\,km\,s$^{-1}$ 
(Sect.~\ref{anal}). The Cartesian velocity components were calculated  
from  $|V_s|$ assuming isotropy.  

The simulation begins with  spatially coincident cluster and bulge spheres,
placed to orbit around   a massive particle 
of $(7\times10^9$\,M$_\odot)$, representing the bulge mass internal to 
the cluster position  (Sect.~\ref{anal}).
The status of the simulation after 10\,Myr is  shown in Figs.~\ref{fig:simul}c,d, 
respectively  for the cluster and  
nine bulge particles  (equivalent to
90\,M$_\odot$) which were captured. In order to further check the capture 
stability we left the simulation evolve up to 50\,Myr 
(Figs.~\ref{fig:simul}c,d) and seven particles (70\,M$_\odot$) remained trapped.
We remind that during these supplementary 40\,Myr no new encounters with bulge
stars were considered.  The capture rate obtained from this numerical
simulation  implies that about 7000 to 9000\,M$_\odot$ can be captured
by a   10$^5$\,M$_\odot$ globular cluster lurking in the bulge during 1\,Gyr, 
in good agreement with the analytical results of Sect.~\ref{anal}. 

\section{Discussion and concluding remarks}
\label{conclu}

In the present {\it Letter} we proposed a  capture mechanism to 
explain a metal-rich secondary RGB in the CMD of the bulge metal-poor
globular cluster HP\,1. So far the proposed mechanisms affecting the number 
of stars in globular clusters are the escape of stars by evaporation (e.g. 
Spitzer \& Thuan 1972)  and ejection  (e.g H\'enon 1969). 
By far evaporation
is the most important escape mechanism: typically, a globular
cluster with 10$^6$\,M$_\odot$ should be stable during a Hubble time, 
whereas
one with 10$^5$\,M$_\odot$ should lose an important fraction of its mass.
The disk shocking process (Ostriker \etal 1972; see also Binney \& 
Tremaine 1987), not only by the disk but also by the bulge itself 
in such central regions, is another possible loss mechanism.  
The presently proposed accretion mechanism for bulge 
globular
clusters would balance such losses, allowing those with
$\sim10^5$\,M$_\odot$ to survive in the bulge. 

\begin{figure}[ht]
\vspace{8 cm}
\includegraphics{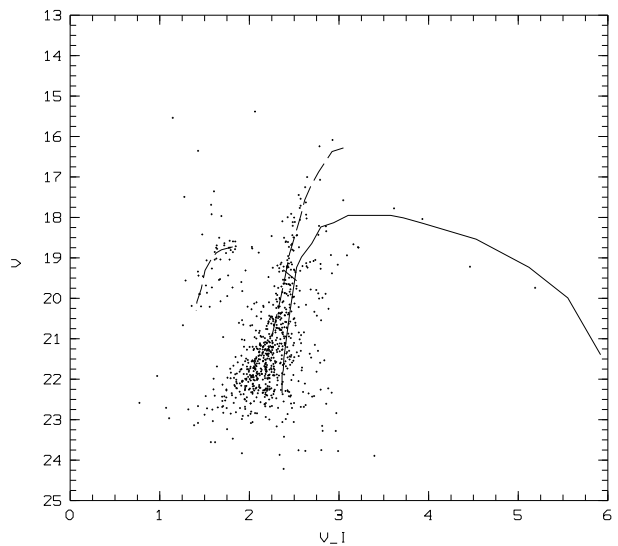}
\caption{Color-magnitude diagram of HP\,1 from OBB97, where in addition to
the mean locus of NGC\,6752 (dashed line), we superimposed that of
NGC\,6553 (continuous line) which  fits the metal-rich sequences}
\label{fig:cmd}
\end{figure}                                                     

Depending on the cluster mass, as well as on the evaporation and capture  
rates it is possible to envisage  changes of the cluster stellar 
population content over a Hubble time. A very massive cluster, initially 
metal-poor,  
would conserve this character and add a secondary
metal-rich component to its CMD, as could be the case of HP\,1.  
On the other hand, a less massive initially metal-poor cluster might recycle
its stellar content, if  capture  dominates over evaporation effects,
so that its CMD would become similar to that of the bulge. In this case the 
metal-poor 
evolutionary sequences like blue HB and vertical RGB would become relatively
less populated.  Hubble Space Telescope color-magnitude diagrams of metal-rich 
bulge clusters would be important to check this scenario. 

The metallicity distribution of the galactic globular clusters is skewed 
towards high metallicities, or it is perhaps 
bimodal (Zinn 1980), and cannot be described by a one-zone metal 
enrichment model 
(Bica \& Pastoriza 1983). The above described scenario of bulge star 
captures by globular clusters and recycling of their stellar population 
provides a natural way
to create an asymmetrical histogram of cluster metallicities from an initially
Gaussian distribution.  It is possible to speculate the existence of two 
families of metallic clusters: (i) genuine ones formed from enriched gas,
and (ii) recycled ones which were initially metal-poor and, over  a Hubble
time, developed CMD sequences like those of the Baade window by means of 
captures.

The relative loci of evolutionary sequences in the HR diagram for globular 
clusters of different metallicities
is fundamental for the calibration of stellar population parameters. Da Costa
\& Armandroff (1990) studied  metal-poor globular clusters together with 
47\,Tucanae, whereas Bica, Barbuy \& Ortolani (1991) studied those 
of nearly solar metallicity.
The HP\,1 CMD containing  combined metal-poor and metal-rich 
stellar populations could provide a means to directly measure relative 
blanketings,  independently of  stellar atmosphere and spectral models.
We show in  Fig.~\ref{fig:cmd}  the CMD of  the spatial 
extraction for r$\leq$23'' from OBB97, where in addition to the mean 
locus of a metal-poor globular cluster 
(NGC\,6752 with [Z/Z$_\odot$]\,=\,-1.54) fitted to the sequences of 
HP\,1 itself, we also superimpose that of the nearly-solar metallicity
globular cluster NGC\,6553 (Ortolani et al. 1995). This HP1 CMD was built from 
images obtained at the ESO NTT 3.55\,m telescope and was reduced with the 
DAOPHOT II package with particular care for crowded field extractions and 
calibrations (Ortolani \etal 1996, OBB97).

We illustrate relative measures of reference points in the CMD. The magnitude 
difference between the metal-poor and metal-rich horizontal branches is 
$\Delta V_{HB} \approx 0.55$\,mag, which has important implications for the 
relative distances of globular clusters (e.g. Jones \etal 1992). 
The magnitude difference between 
the brightest giants in the metal-poor and metal-rich RGBs 
is $\Delta V_{BG} \approx 1.75$\,mag. However, if the tips of the metal-poor 
and metal-rich giant branches correspond to a similar stellar temperature,
the difference could be as large as $\Delta V_{TIP} \approx 3.6$\,mag,  due to 
blanketing effects. Finally, the color difference could be as large as 
$\Delta (V-I)_{TIP} \approx 2.2$\,mag. Such magnitude and color differences 
are fundamental quantities for blanketing calibrations of metal-rich 
stellar populations.

\acknowledgements

This work was partially supported by the Brazilian institutions CNPq and FAPESP.
The numerical calculations were made in the National Supercomputing Center
(CESUP/RS), operated by Universidade Federal do Rio Grande do Sul (UFRGS). 


\newpage

\end{document}